\documentclass{aa}

\usepackage{graphicx}
\usepackage{txfonts}
\usepackage{xcolor}
\definecolor{myblue}{rgb}{0.00, 0.0, 0.9}
\definecolor{myred}{rgb}{0.90, 0.0, 0.0}
\definecolor{mygreen}{rgb}{0.0, 0.7, 0.0}
\usepackage[breaklinks,  citecolor=myblue, linkcolor=myred, urlcolor=purple, colorlinks=true, debug, baseurl=' ']{hyperref}
\def \HI{H\,{\sc i}} % this is for HI notation

\begin{document}

\title{The magnetic field of the Radcliffe Wave:  starlight polarization at nearest approach to the Sun}

\author{
G.~V.~Panopoulou,\inst{1}\fnmsep \thanks{E-mail: georgia.panopoulou@chalmers.se}
C.~Zucker,\inst{2}
D.~Clemens\inst{3},
V.~Pelgrims\inst{4},
J.~D.~Soler\inst{5},
S.~E.~Clark\inst{6,7},
J.~Alves\inst{8},
A.~Goodman\inst{2},
J.~Becker Tjus\inst{1,9,10}
}

% List of institutions
\institute{$^1$ Department of Space, Earth and Environment, Chalmers University of Technology, Gothenburg, Sweden\\
$^2$ Center for Astrophysics $\vert$ Harvard \& Smithsonian, 60 Garden St., Cambridge, MA 02138, USA\\
$^3$ Institute for Astrophysical Research, Boston University, 725 Commonwealth Avenue, Boston, MA 02215, USA\\
$^4$ Universit\'{e} Libre de Bruxelles, Science Faculty CP230, B-1050 Brussels, Belgium\\
$^5$ Istituto di Astrofisica e Planetologia Spaziali (IAPS), INAF, Via Fosso del Cavaliere 100, I-00133 Roma, Italy\\
$^6$ Department of Physics, Stanford University, Stanford, CA 94305, USA\\
$^7$ Kavli Institute for Particle Astrophysics \& Cosmology, Stanford University, P.O. Box 2450, Stanford, CA 94305, USA\\
$^8$ University of Vienna, Department of Astrophysics, T\"urkenschanzstrasse 17, 1180 Vienna, Austria\\
$^9$ Theoretische Physik IV, Fakultät für Physik \& Astronomie, Ruhr-Universität Bochum, 44780 Bochum, Germany\\
$^{10}$ Ruhr Astroparticle and Plasma Physics Center (RAPP Center), Ruhr-Universität Bochum, 44780 Bochum, Germany
}

% These dates will be filled out by the publisher
\date{Accepted XXX. Received YYY; in original form ZZZ}

% Enter the current year, for the copyright statements etc.

% Don't change these lines

\abstract
{}{We investigate the geometry of the magnetic field towards the Radcliffe Wave, a coherent 3-kpc-long part of the nearby Local Arm recently discovered via three-dimensional dust mapping.}{We use archival stellar polarization in the optical and new measurements in the near-infrared to trace the magnetic field as projected on the plane of the sky. Our new observations cover the portion of the structure that is closest to the Sun, between Galactic longitudes of \,122$^\circ$  and  188$^\circ$.}{The polarization angles of stars immediately background to the Radcliffe Wave appear to be aligned with the structure as projected on the plane of the sky. The observed magnetic field configuration is inclined with respect to the Galactic disk at an angle of 18$^\circ$. This departure from a geometry parallel to the plane of the Galaxy is contrary to previous constraints from more distant stars and polarized dust emission. We confirm that the polarization angle of stars at larger distances shows a mean orientation parallel to the Galactic disk.}{We discuss implications of the observed morphology of the magnetic field for models of the large-scale Galactic magnetic field, as well as formation scenarios for the Radcliffe Wave itself.}

\keywords
{ISM: magnetic fields -- polarization -- ISM: structure -- local interstellar matter -- dust, extinction
}

\titlerunning{The magnetic field of the Radcliffe Wave}
\authorrunning{Panopoulou, G.~V. et al.}

\maketitle

\section{Introduction}
\label{sec:introduction}

The magnetic field is one of the most elusive components of the Milky Way, mainly because of the difficulties
associated with measuring it  \citep{Beck2013pss5.book..641B,Haverkorn2015,Han2017}. Our understanding of the magnetic field is drastically improving, partly due to the
advent of the \textit{Planck} mission \citep{Planck_mission2011}. Observations of the polarization of thermal dust emission have revealed the
morphology of the magnetic field in the dusty interstellar medium (ISM) with unprecedented sky coverage \citep{planck2020XII}.

On scales of hundreds of parsecs to roughly one kiloparsec, the \textit{Planck} data in the Galactic plane trace a magnetic field that is
parallel to the disk \citep{PlanckXIX2015_dust,Planck2015XXI}, confirming what had been known from starlight polarization \citep{Mathewson1970, heiles2000,Fosalba2002}. However, new results from the Galactic Plane Infrared Polarization Survey \citep[GPIPS;][]{Clemens2020} find significant variations among the polarization angles of stars, with regions showing departures from the disk orientation within the inner Galaxy. It remains unclear what physical scales these variations of the cumulative polarization of stars are probing, potentially arising, for example, from the presence of dense molecular clouds.
On scales of tens of parsecs, the magnetic field orientation can vary substantially from the disk geometry when probing dense clouds
\citep[e.g.][]{Stephens2011,Marchwinski2012,Planck2016_XXXV_clouds}. On such scales, the \textit{Planck} data show that the magnetic field
orientation is correlated with that of dust structures \citep{PlanckXXXII2016} and linear features in the neutral atomic hydrogen (H{\sc i}) emission throughout the sky \citep{Clark2014, Clark2015}. The origin of this correlation has been connected to the properties of magneto-hydrodynamic (MHD)
turbulence \citep[e.g.][]{Soler2017,Xu2019}. However further observational evidence is needed to fully understand the coupling of the magnetic field and density across scales \citep[e.g.][]{ClarkPeek2019, Hennebelle2019}.  

These advances in mapping the magnetic field have coincided with significant improvements in our ability to reconstruct the three-dimensional (3D) distribution of dust (c.f. \citealt{Green_2019}, \citealt{Lallement_2022}, \citealt{Edenhofer}), thanks to the advent of the \textit{Gaia} mission \citep{Gaia2016}. 
These 3D dust maps are transforming our view of the ISM structure within a few kiloparsecs from the Sun, revealing new and unexpected structures in the 3D density distribution \citep{Zucker2023}. 
One of the most striking new discoveries is the Radcliffe Wave (RW) --- a 2.7-kpc-long structure with an aspect ratio of roughly $20:1$, which also hosts many nearby star-forming regions \citep{Alves2020}. 

The RW seems to be a prominent feature of Galactic structure, argued (e.g., \citealt{Swiggum_2022}) to be the gaseous reservoir of the Local Arm \citep{Reid_2019} in the solar vicinity. It exhibits the puzzling shape of a damped sinusoid extending above and below the midplane of the Galaxy with an amplitude of roughly 160~pc and crossing the midplane near Galactic longitude $l = 165^\circ$. 
Despite its prominence, there remain important open questions regarding its origin and role in the history of the local ISM. It is possible that the wave was caused by a perturber that collided with the disk \citep{Thula_2022}, though internal mechanisms, including a series of supernova explosions that displaced the gas from the midplane, are also possible \citep{Tu_2022}. Using new constraints on the 3D space motions of young stellar clusters detected in the wave with Gaia DR3, \citet{Konietzka2024} show that the structure is oscillating with a maximum vertical velocity (perpendicular to the disk of the Milky Way) of $v_z \rm \approx 14 \; km \; s^{-1}$ \citep[see also][]{Li_2022}.

The existence of this feature is perplexing in terms of our understanding of the Galactic Magnetic Field (GMF). Measurements of stellar polarization have been used to determine the mean direction of the magnetic field in the Local Arm \citep{Heiles1996}, finding that the field runs parallel to the Galactic plane. 
At the same time, the RW is part of the Galactic disk but does not lie parallel to the disk: it appears to undulate above and below the disk. This apparent discrepancy between the orientation of the magnetic field and the shape of the RW calls for a detailed investigation.

In this paper, we performed a study of the magnetic field towards the RW. Our aim is to trace the Galactic magnetic field in the vicinity of the RW and determine whether it has been affected by the presence of the RW. We use starlight polarization in combination with stellar distances to probe the magnetic
field morphology at the distance to the RW. 
Section~\ref{sec:data} presents the data used in this study. 
Section~\ref{sec:methods} describes the statistical treatment of the stellar polarization data. Section~\ref{sec:results} compares the magnetic field geometry as traced by stellar polarimetry to the morphology of the RW, and shows that within 400~pc of the Sun, the mean magnetic field is preferentially aligned with the RW and not with the Galactic plane at longitudes $l=122 - 188^\circ$. Our findings are discussed in Sect.~\ref{sec:discussion} and conclusions are provided in Sect.~\ref{sec:conclusions}.

%%%%%%%%%%%%%%%%%%%%%%%%%%%%%%%%%%%%%%%%%%%%%%%%%%

\section{Data}
\label{sec:data}

\subsection{3D dust extinction map}

We use the publicly available 3D dust extinction map from \citet{Edenhofer} to trace the distribution of dust towards the RW. The map is constructed using  54 million stars from \citet{Zhang2023}, who forward-modeled the stars' atmospheric parameters, distances, and extinctions using the low-resolution Gaia BP/RP spectra \citep{Carrasco_2021}. We choose the \citet{Edenhofer} map because it achieves good spatial resolution both on the plane-of-sky (POS) and along the line-of-sight (LOS), with 14\arcmin ~angular resolution and parsec-scale distance resolution. The \citet{Edenhofer} map extends out to 1.25~kpc from the Sun, which encompasses the bulk of the RW. 
We use the publicly available version of the map\footnote{\url{https://doi.org/10.5281/zenodo.8187943}} provided in HEALPix format \citep{Gorski2005}.
This is a collection of 516 HEALPix maps of $N_{\rm{side}}=256$, each corresponding to a different logarithmically-spaced distance bin along the LOS, spanning distances from 69 to 1244~pc. We use the ``mean'' value in each pixel, which is given in arbitrary units of differential extinction.
Following the recommendation from \citet{Edenhofer}, we multiply the values of the map by 2.8 to obtain $A_{\rm{V}}$ in magnitudes, based on the published extinction curve from \citet{Zhang2023}.

\begin{figure*}
\centering
\includegraphics[scale=0.7]{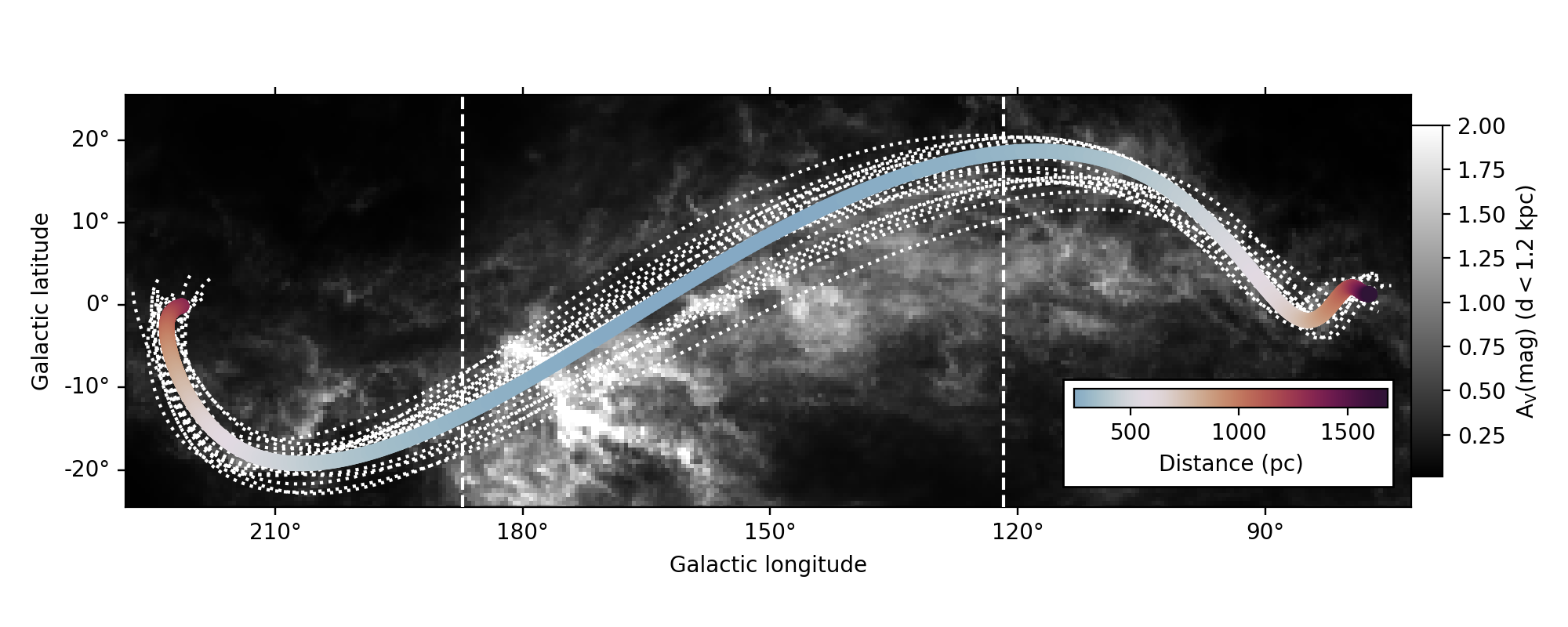}
\caption{The default model of the RW spine (thick coloured line) as projected on the sky. Colours indicate the distance of the RW from the Sun at the corresponding point along the RW spine. The background image shows the extinction integrated out to 1.25%1.2
kpc from \citet{Edenhofer}. The 20 alternative RW models are shown as dotted white lines. Dashed vertical white lines mark the span of the longitude range of interest. }
\label{fig:RWshape}
\end{figure*}

\subsection{Radcliffe Wave model}
\citet{Alves2020} constrain the ``spine'' of the RW in Heliocentric Galactic cartesian space by fitting a damped sinusoidal model to the 3D distribution of nearby molecular clouds from \citet{Zucker2020}. Their model for the RW spine was obtained using a dynamic nested
sampling package for estimating Bayesian posteriors and evidences \citep{Speagle2020}. We use their best-fit model\footnote{\url{https://doi.org/10.7910/DVN/OE51SZ}} as our default model, and also use 20 samples drawn from the posterior distribution of the solution to investigate the dependence of our results on the choice of model. We refer to the 20 samples as `alternative' models in the rest of the text.
We convert each RW spine model to a spherical coordinate system of longitude, latitude, and distance ($l,b,d$) to compare with the polarization measurements both on the plane of the sky and along the line of sight. 

Figure \ref{fig:RWshape} shows the default model of the RW spine as projected on the sky. Points on the spine are coloured according to their distance from the Sun. The background shows the $A_{\rm V}$ map from \citealt{Edenhofer}, integrated out to a maximum distance of 1.25\,kpc. The model extends over a large range of longitudes: $l = (78^\circ,224^\circ)$ and intersects the midplane ($b = 0^\circ$) at $l= 164^\circ$. The 20 alternative models are also shown as white dotted curves. They span a range of locations and shapes.

\subsubsection{Polarized dust emission from \textit{Planck}}

We use the {\it Planck} 353~GHz maps of Stokes $Q$ and $U$ to study the polarization angles of dust emission in the sky region containing the RW. 
We select the 80\arcmin-resolution maps produced via the Generalized Needlet Internal Linear Combination (GNILC) algorithm \citep{Remazeilles2011}, which reduces the contamination from the Cosmic Microwave Background (CMB) and instrumental noise in polarization \citep{Planck2018_IV}.  
We smooth the Stokes $Q$ and $U$ maps and their covariance matrices to a FWHM = 2$^\circ$ using the \texttt{smoothing} function of the {\tt healpy} package and the procedure described in Appendix A of \citep{PIP_XIX}. 

The polarization angle of the dust emission in the Galactic reference frame according to the IAU convention is:
\begin{equation}
    \phi_{\rm{dust}} = 0.5\,{\rm{arctan}}(-U,Q)
\end{equation}
where we have used the 2-argument arctangent function. We rotate the angles by 90$^\circ$ to obtain the corresponding plane-of-sky magnetic field orientations, $\theta_{dust}$. The polarized intensity and its uncertainty are computed as:
\begin{equation}
    P = \sqrt{Q^2 + U^2}, \,\,\,\,\, \sigma_P = 1/P \sqrt{Q^2 C_{QQ}+ U^2 C_{UU}},
\end{equation}
where $C_{QQ}, C_{UU}$ are the diagonal terms of the covariance matrix and we have ignored correlations between $Q$ and $U$ (see also equation B.4, \citealt{PIP_XIX}). 
We calculate the uncertainty in polarization angle as:
\begin{equation}
    \sigma_{\phi,dust} = 28.65^\circ \sigma_P/P,
\end{equation} 
which is a good estimator of the uncertainty at high S/N \citep{Naghizadeh-Khouei1993}.
%which is a good approximation for high S/N ($P/\sigma_P > 4$).

\subsection{Starlight polarization data}
\label{sec:stars}

We use a combination of archival data and new, targeted observations to obtain a sample of stars with stellar polarization at known distances probing the RW. These datasets are described below.

\subsubsection{Compilation of optical polarization catalogs from the literature}

We use the compilation of stellar polarization catalogs presented in \citet{Panopoulou2023} (hereafter P2023). This compilation combines optical polarimetry for $\sim$~55,000 stars from a large body of published literature. We use the data from their Table~6, which contains polarimetry and distances from \textit{Gaia} EDR3 for $\sim$~42,000 stars. We remove sources flagged as intrinsically polarized. We also remove stars with unknown uncertainties in the polarization fraction and polarization angle, resulting in a catalog of 35,864 stars over the sky. We make further selections based on the stars' positional proximity to the RW in Sect. \ref{sec:methods}.

\subsubsection{New NIR polarization data from Mimir}
\label{sec:mimir}

\begin{figure*}
\centering
\includegraphics[scale=0.9]{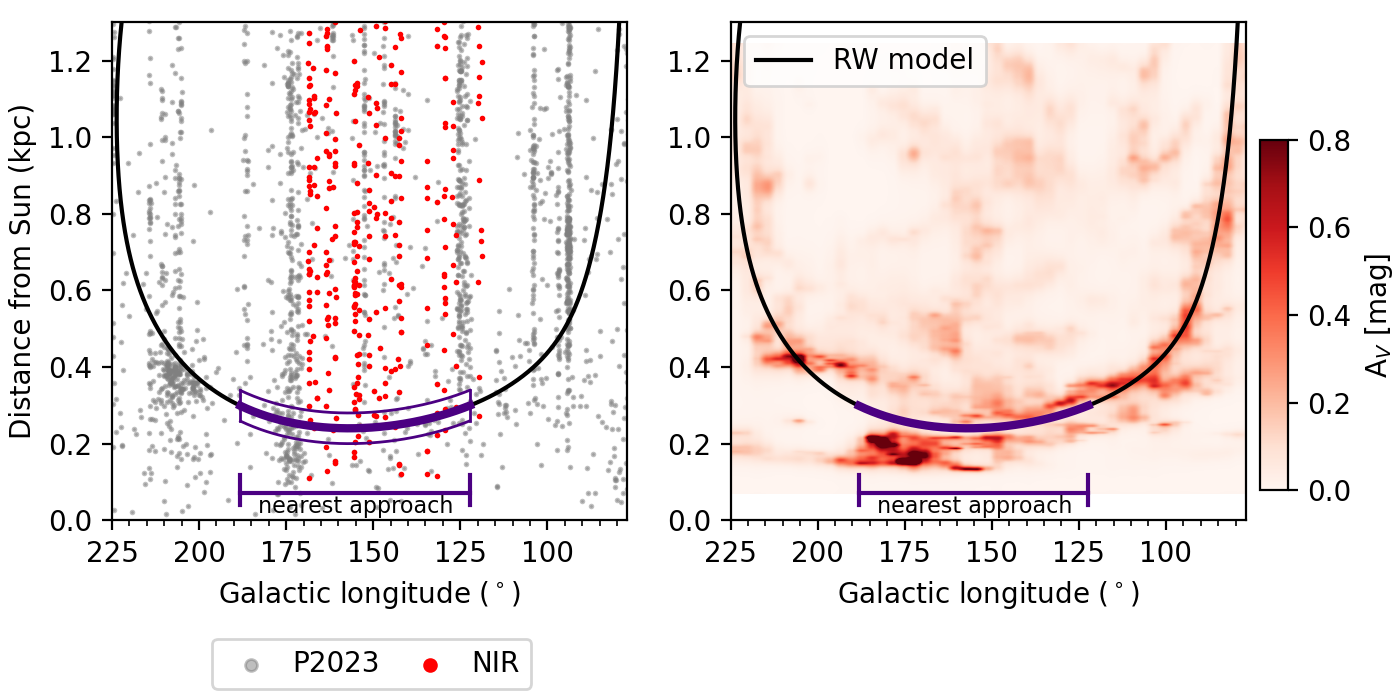}
\caption{Selection criteria in the distance-longitude diagram. Left: Default RW spine model (black line) with nearest-approach longitude range marked in purple. The purple outline marks the part of the RW spine included in the nearest approach longitude range. Filled circles mark the positions of stars in the polarization catalogs we employ, namely the optical P2023 compilation (gray), %(black), 
and the two samples in the NIR from Mimir (red). Only stars with high-significance measurements of distance and polarization are shown. Right: Extinction integrated along the latitude axis, from \citet{Edenhofer} for pixels within 10$^\circ$ of the default RW spine.} 
\label{fig:approach}
\end{figure*}

We conducted a targeted survey of stellar polarization along the nearby portion of the RW  using the Mimir NIR polarimeter \citep{Clemens2007}. 
To ensure a measurable polarization signal, we selected fields with $A_{\rm{V}} > $ 1.4 mag based on the 3D dust map of \citet{Green_2019}, integrated out to 350\,pc (covering the nearest distance of the RW, see Fig.~\ref{fig:approach}). While we utilize the \citet{Edenhofer} 3D dust map for the majority of the analysis, we originally chose the \citet{Green_2019} for target selection, as it was the highest-angular resolution 3D dust map available in the literature at the time of observations. A similar 3D dust mapping methodology used by \citet{Green_2019} was used in \citet{Alves2020} to originally detect the RW (see \citealt{ZuckerSpeagle2019}; \citealt{Zucker2020}). The fields were selected to lie within $\sim 5^\circ$ of the RW spine and in areas that
did not have existing stellar polarization measurements from the literature. By cross-matching the 2MASS catalog \citep{Skrutskie2006} with \textit{Gaia}, we further required the observed regions to have at least 4 stars each within the 10\arcmin $\times$ 10\arcmin ~field of view of Mimir, at distances  $d\leq350$~pc and that were bright enough to have significant detection of the polarization (apparent H-band magnitude $m_H<$13.5 mag). 

For each pointing, we rotated the half-wave plate to 16 different position angles, with fixed %10~s  
 integration times for each position. Sky dithering was performed in six positions, with offsets of typically 15 arcsec.
This resulted in 6$\times$16=96 images per observation. Following \citet{Pavel2011}, we observed each field using multiple exposures: a short exposure of 2.3~s and two long exposures of 15~s per image.
Observations were conducted during January/February 2020 and January 2022. %We note that a hardware failure was identified after the initial run of observations --- due to low temperatures the half-wave-plate position determination switch was not responding correctly. The malfunction was later corrected by heating the relevant parts when the ambient temperature dropped below a critical point. As a result, a set of our initial observations were compromised by systematic effects that we could not correct for. These fields were re-observed in 2022.
The final catalog of stellar polarizations contains measurements towards 19 fields across the length of the RW, within the longitude range $l$ = [122$^\circ$, 188$^\circ$].

The data reduction was done with the IDL software packages described in \citet{Clemens2012}. 
%The two long observations were 'meta-grouped' to produce a final image for each RW field. 
The reduction was performed separately on each series of short and long exposures, producing three polarization catalogs. The catalogs contain information on the relative Stokes parameters $q = Q/I,\, u=U/I$ (where $I$ is the total intensity of the star), their uncertainties, as well as stellar coordinates, star identifiers and photometry from 2MASS. 
These polarization catalogs for the short and two long exposures were merged by matching the common stars and computing the weighted average Stokes $q$ and $u$ parameters based on their corresponding uncertainties.  The fractional linear polarization, $p$, and Electric Vector Position Angle (EVPA), $\chi$, are defined from the Stokes parameters as:
\begin{equation}
    p = \sqrt{q^2 + u^2}, \,\,\,\,\,\,\,\,
    \chi = 0.5 \,{\rm{arctan}}(u,q),
\end{equation}
where the 2-argument arctangent function is used. In this work, we do not correct for bias in $p$ \citep{Vaillancourt2006}, as we are interested solely in the polarization angle (and its uncertainty).

We applied quality cuts on the output polarization catalog, keeping stars with H-band brightness 
$ m_H\leq 13 \; \rm mag$ and signal-to-noise (S/N) in the biased polarization fraction of 
$p/\sigma_p \geq 2$.
Stars not satisfying these criteria were rejected from the final catalog. In total 477 stars passed these selection criteria and were included in the final catalog. These data will be made publicly available on CDS \url{https://cdsweb.u-strasbg.fr/} and on the online version of the journal.

\subsubsection{NIR polarization data of open clusters from Mimir}

\citet{Hoq2015} and \citet{Hoq2017PhDT} reported Mimir H-band polarimetry obtained toward fields containing 31 Open Clusters in the outer Milky Way. 
The data collection mode was similar to that described in Sect. \ref{sec:mimir}, but with integration times chosen to match stellar brightnesses in each cluster. As such, the limiting magnitude varies for each observation. 
The 14 clusters, spanning $\ell$ = 119$^\circ$ -- 168$^\circ$, which contributed data to this current study included Berkeley~12, Berkeley~14, Berkeley~18, Berkeley~60, Berkeley~70, King~1, King~5, King~7, NGC~559, NGC~663, NGC~869, NGC~1245, NGC~1857 and NGC~2126. 
The dates of observations range from January 2006 to January 2013. The limiting polarimetric magnitude, which accounts for 90\% of all stars brighter than that value, ranges from 11.1 to 16.4 across the 14 cluster sample. 
Although the distances to the clusters range from 1.0 to 6.2\,kpc \citep{Hoq2015}, all stars in each field were tested for \textit{Gaia} matches and other selection effects. 
The same quality criteria were applied as for stars in the RW survey ($ m_H \leq 13$~mag and  $p/\sigma_p \geq 2$). In total, 893 stars from the 14 selected clusters met all selection and \textit{Gaia}-match criteria and were used in the analysis.

\subsubsection{Data handling and cross-match with Gaia}

We cross-matched the Mimir polarization catalogs (RW survey and Open cluster surveys) with \textit{Gaia} DR3 using a search radius of 1 arcsecond. For the 12 sources that returned multiple matches, we selected the brightest source among the matches. The final catalog from the Mimir data contains 1371 unique sources with \textit{Gaia} matches. To obtain stellar distances, we use the latest \textit{Gaia}-based catalog providing probabilistic distance estimates based on parallax measurements  \citep{Bailer-Jones2021}. We cross-matched our catalog with that of \citet{Bailer-Jones2021} based on the \textit{Gaia} source identifier (which is the same for EDR3 and DR3). Throughout this work, we use the photo-geometric distance estimates from the \citet{Bailer-Jones2021} catalog. The P2023 catalog provides \textit{Gaia} EDR3 matches and stellar distances from \citet{Bailer-Jones2021}.

We convert the polarization angles $\chi$ measured in the celestial frame (according to the IAU convention, increasing from North to the East) to angles in the Galactic reference frame, $\theta$, through \citep[e.g.][]{Appenzeller1968}:
\begin{equation}
    \theta = \chi + \arctan ( \frac{\sin(l_{NCP}-l)}{\tan b_{NCP} \cos b - \sin b \cos (l_{NCP} - l) } )
\end{equation}
where $l_{NCP}, b_{NCP}$ are the Galactic longitude and latitude of the North Celestial Pole and $l, b$ are the Galactic coordinates of each star.

\section{Methods}
\label{sec:methods}

We aim to determine whether the Galactic magnetic field geometry shows a disturbance associated with the undulating pattern seen in the dust structure that defines the RW. We use stellar polarization to trace the morphology of the magnetic field as projected on the sky. By selecting stars whose light is primarily extinguished and polarized by the RW, we can probe the plane-of-sky component of the magnetic field that aligns dust grains in the RW. 

The flow dynamics leading to the existing undulating morphology of the RW likely have disturbed the magnetic field. To reject the hypothesis that the magnetic field is parallel to the midplane over the extent of the RW, we would need to detect a region with a magnetic field orientation that significantly departs from plane-parallel. We focus our analysis on the nearest portion of the RW. This choice simplifies the analysis of the magnetic field in two ways. First, it lifts the need for tomographic decomposition to trace the magnetic field, as would be needed if multiple components along the LOS contributed to the stellar polarization signal. If the RW is the first polarizing screen along the LOS, then we can simply trace its magnetic field by measuring the polarization of stars immediately background to it. Second, the analysis of \citet{Heiles1996} shows that the starlight polarization fraction is greater in the longitude range of interest, compared to other directions along the Local Arm. This implies that polarization in this area will be more easily detected. 

\subsection{Region of nearest approach}

The RW spans a large range in Galactic longitude and distances from the Sun, while remaining within a smaller range of latitude, as shown in Fig.~\ref{fig:RWshape}.  
We define the region for our study as the portion of the RW within distance $d_{RW} < 300 \; \rm pc$ from the Sun. This distance cut corresponds to longitudes $l = [122^\circ,\, 188^\circ]$. 
We additionally impose a latitude cut of $|b| < 25^\circ$, and restrict our analysis to sightlines within 10$^\circ$ of the RW spine, encompassing the bulk of the extinction integrated along the LOS out to the limits of the \citet{Edenhofer} map.
We present the spine of the RW in the longitude-distance plane in Fig.~\ref{fig:approach} (left). 
Stars in our catalogs are shown as dots in the figure, with different colors specifying the different surveys. The purple region marked on the spine of the RW denotes the longitude range where the RW is within 300 pc. We refer to this region of interest as the ``nearest approach'' -- the area where the RW reaches its smallest distance from the Sun. Within this longitude range, the RW appears to cross the Galactic midplane $(b=0^\circ)$  
forming an angle of $\sim$ 30$^\circ$ with respect to the $(b=0^\circ)$ line near $l = 164^\circ$ (Fig. \ref{fig:RWshape}). Similarly to the best-fit model, the alternative spine models form angles of $25-34^\circ$ with the midplane. %at a projected plane-of-sky angle of $\sim$ 30$^\circ$. 

We calculate the angle between the tangent to the RW spine at point $i$ and the plane of the sky, $\gamma_i$, for each location along the RW. The RW spine in the nearest approach region lies mostly in the plane of the sky, with $\cos^2(\gamma_i) > 0.6$.
The linear extent of the RW model within this longitude range is 350~pc (the separation between the two end-points within the nearest approach region, measured in cartesian coordinates). The height (vertical to the disk, assumed to be at $z = 0$) difference between the two end-points of the model in this region is 160~pc.

\subsection{Star sample selection} \label{sec:stellar_selection}

We wish to select stars whose polarization is primarily due to the RW. Since there is no prior information on the polarization properties of the RW, we investigate the 3D distribution of extinction of the structure. If the extinction towards a star is dominated by dust associated with the RW, then it is likely that the star's polarization too will be dominated by the RW (as long as the magnetic field is not directed along the LOS, in which case negligible polarization would arise). 

We construct a map of the extinction in the distance-longitude plane. Because the range of latitudes is much smaller than the range of longitudes spanned by the RW (Fig. \ref{fig:RWshape}), collapsing along the latitude axis still allows us to view the two main axes of the object (l, d) and provides a simple way to visualize the dust distribution along the lines of sight towards the RW. Since we are interested in visualizing the bulk of the extinction, we downgrade the resolution of each HEALPix map to $N_{\rm{side}} = 64$. For all sky pixels within $10^\circ$ of the RW we extract profiles of the differential extinction as a function of distance using the 3D dust map of \citet{Edenhofer}. 
%In practice, we use the centers of all $N_{\rm{side}} = 64$ pixels within $10^\circ$ of the RW spine. 
We now have independent profiles of differential extinction as a function of distance for the aforementioned sightlines. Since the differential extinction data are sampled on an irregular distance grid, we interpolate each profile to obtain a re-gridded profile sampled at regular distance intervals spaced by 1~pc, the approximate resolution of the \citet{Edenhofer} map within a few hundred parsecs of the Sun. We create a coarse longitude grid with $3^\circ$ spacing. Next, we construct a 2D map of extinction in the longitude-distance plane by summing the profiles of all sightlines within a given longitude bin at each distance. This results in a map of the total extinction as viewed along the latitude axis (spanning the selected latitudes and perpendicular to the longitude-distance plane).

This 2D extinction map is shown in Fig.~\ref{fig:approach} (right). The distribution of $A_{\rm{V}}$ shows overdensities that trace the RW model\footnote{Hereafter, `RW model' refers to the default model unless otherwise specified.} (black line) for the extent of the RW at longitudes $l \lesssim 150^\circ$. Towards $l = 160^\circ - 185^\circ$ there is an offset between the peak of the dust distribution and the RW model. This longitude range encompasses the Taurus Molecular Cloud (TMC). There is also a notable absence of dust along the RW model for longitudes $l > 215^\circ$, beyond the Orion Molecular Cloud (at 0.45~kpc distance). The RW model was constructed by fitting a damped sinusoid function to the locations of discrete molecular clouds \citep{Alves2020}. Consequently, the fact that we do not find the model able to match all of the details of the 3D dust distribution is not surprising. For our purposes, it appears that this model is an adequate description of the large-scale geometry of the dust distribution.

Dust associated with the RW appears to provide the bulk of the extinction for most of our selected sightlines out to the boundaries of the 3D map. However, this is not the case in the longitude range $l \in [100^\circ,$\, 170$^\circ]$, where dust reddening shows overdensities at distances $>$  400~pc that do not appear to be part of the RW (the differential reddening drops to zero in between the RW and those structures). To be conservative, we conclude that the extinction of stars at distances between the RW and 400 pc, within the longitude range $l \in [122^\circ, 188^\circ]$ and within the sky area of 10$^\circ$ from the RW spine, is dominated by the dust associated with the RW. This also suggests that the polarization of those stars will be dominated by the RW, barring 3D magnetic field geometry effects (inclination, LOS tangling within the RW itself that may cause depolarization).

In the following, we distinguish between two stellar samples occupying the same region on the sky (within $10^\circ$ of the RW spine, having $l = [122^\circ, 188^\circ]$ and $|b| < 25^\circ$). The ``far'' sample corresponds to stars with distances beyond 1.2~kpc ($d > 1.2 \; \rm kpc$). We define our default near sample to include stars with distances $ d <$ 400 pc. 
In Sect.~\ref{sec:results} we incrementally increase this distance limit to create near samples out to 1.2~kpc. We place a S/N threshold in polarization fraction: $p/\sigma_p \geq 2.5$, which corresponds to an uncertainty in the polarization angle of $\leq 12^\circ$. The main effect of this S/N threshold is to remove stars that are foreground to the RW, where there is too little dust to induce measurable polarization above our survey sensitivity limits.

A final selection cut is implemented to remove stars towards the TMC. The magnetic field in this cloud may not trace the large-scale magnetic field of the RW. In the TMC, the magnetic field has been perturbed by a nearby supernova explosion, forming the so-called ``Per-Tau'' shell, as well as by other smaller-scale feedback events
(see e.g., \citealt{Chapman2011};\citealt{Bialy2021}; \citealt{Doi2021}; \citealt{Soler2023}; \citealt{Konietzka2024}). The cloud appears to be squeezed between the Per-Tau shell and the Local bubble \citep{Zucker2022Natur}. 
We define a circular region centered on $(l\,,b) = (172.6^\circ,\, -15.6^\circ)$, following Table~1 of \citet{Zucker2021_clouds}. We chose a radius of $10^\circ$, which encompasses the entire length of the TMC as found in that work and removed all stars in our sample within that area.

\begin{figure*}
\includegraphics[width=0.9\textwidth]{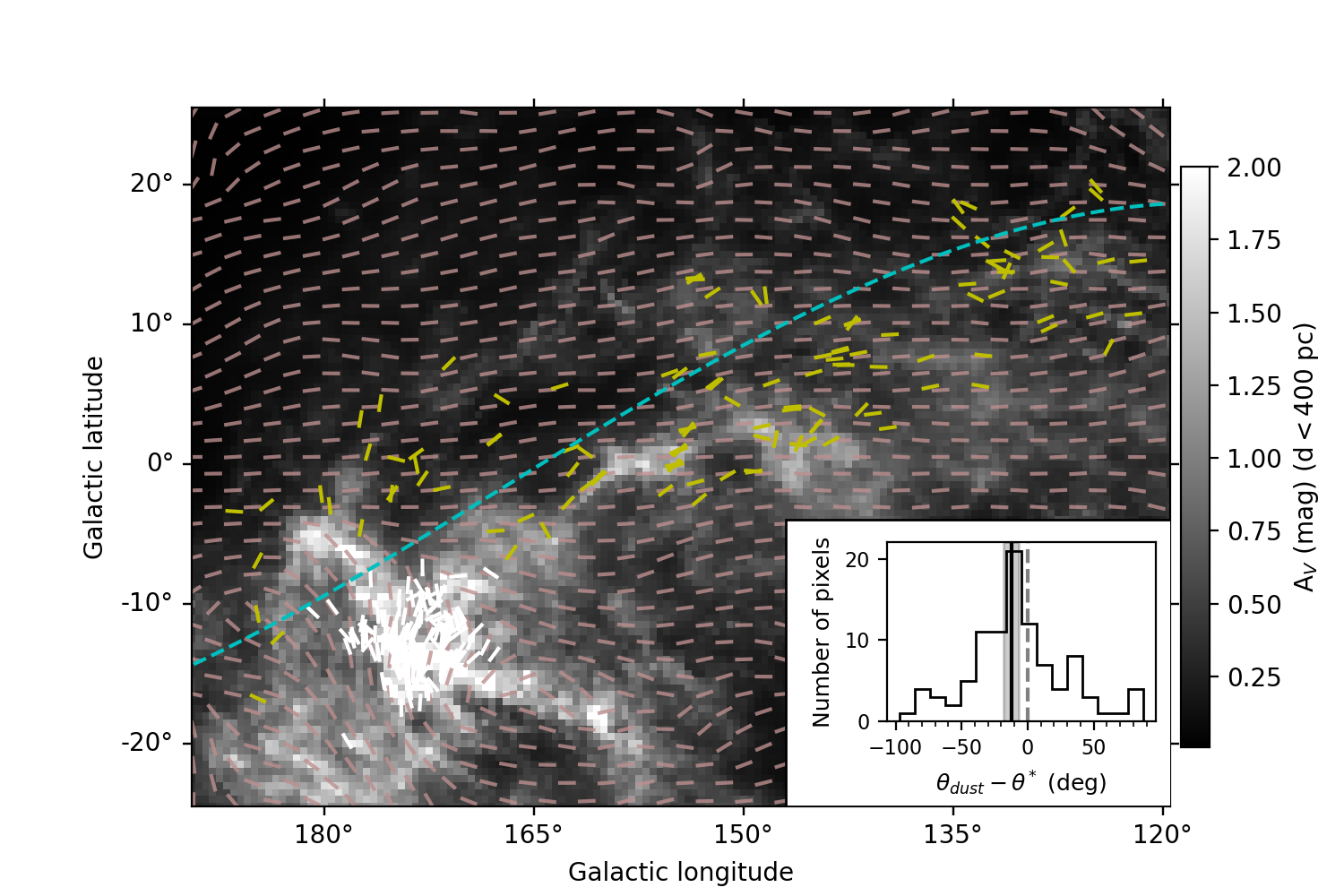}
\caption{Comparison between the pixelated polarization angles of stars and those of dust emission in the nearest approach region of the RW to the Sun. The polarization angles of stars with significant polarization fraction, within 400 pc from the Sun and within 10 degrees angular separation of the RW are shown as yellow segments. White segments show the stars excluded from the analysis due to their projected proximity to the TMC. Light brown segments show the orientation of the plane-of-sky magnetic field derived by rotating the \textit{Planck} 353 GHz data by 90 degrees. The cyan dashed line shows the spine of the RW as projected on the sky, while the background image shows the extinction out to 400 pc. The inset on the bottom right shows the distribution of angle differences between the \textit{Planck}-inferred magnetic field orientation and the polarization angles of stars at $N_{\rm{side}} = 64$ excluding sightlines towards the TMC. The circular mean and its error are shown as a vertical black line and a gray band. The dashed vertical gray line in the inset indicates a relative orientation of $0^\circ$.}
\label{fig:planck_stars}
\end{figure*}

\subsection{Producing pixelized stellar polarization data}
\label{sec:pixelization}

Our aim is to compare the mean orientation of the  GMF with that of the RW on scales larger than those of individual clouds (i.e. $\sim 5-10~\rm pc$, \citealt{Zucker2021_clouds}). At the nearest distance to the RW of 300~pc, these scales correspond to angular separations of 1-2 degrees. We therefore wish to homogenize the sampling over the entire sky area of interest to these scales. The sightlines towards stars with measured polarization are unevenly spaced on the sky. For example, the fields observed with Mimir towards the RW may contain tens of stars at all distances within the $10\arcmin \times 10\arcmin$ field of view. By averaging the stellar polarization data over degree-sized scales we avoid overweighting sky pixels with many stellar measurements as a result of the observing strategy.

We pixelize the stellar polarization angles to $N_{\rm{side}} = 64$. Within each pixel, we calculate the weighted mean polarization angle, $\theta^*$, of the $N$ stars within the pixel weighting by their inverse variances:
\begin{equation}
    \theta^* = \frac{1}{2} \, {\rm{arctan}}\left(  \rm \overline{u} , \overline{q} \right),
    \label{eqn:circmean}
\end{equation}
as appropriate for circular data (e.g., \citealt{Fisher1995}), where we use the two-argument arctangent function. {The quantities $ \rm \overline{u} , \overline{q}$ are: }
\begin{equation}
    {\rm \overline{q}} = \frac{1}{W}\sum_{i=1}^{N} w_i \cos(2\theta_i)
\end{equation}

\begin{equation}
    {\rm \overline{u}} = \frac{1}{W}\sum_{i=1}^{N} w_i \sin(2\theta_i),
\end{equation}
where $W$ is the sum of the weights, $w_i$, with  $w_i = (\sigma_{\theta_i})^{-2}$ and $W = \sum_{i=1}^N w_i$. We wrap all resulting angles to the range [0, $\pi$]. {We compute the error on the weighted mean quantity $\rm \overline{q}$ (and similarly for $\rm \overline{u}$) according to the second method presented in \cite{std_mean}:}
\begin{equation}
\sigma_{\rm \overline{q}} = \sqrt{\frac{N}{(N-1) \sum_{i=1}^N w_i^2} \left[ A + B + C \right]}\label{eq:sigma}
\end{equation}
where:
\begin{equation}
A = \sum_{i=1}^N \left( w_i {\rm q_i} - \overline{w_i {\rm q_i}} \right)^2,
\end{equation}
\begin{equation}
B = 2 \overline{{\rm q_i}} \sum_{i=1}^N \left[ (w_i - \overline{w_i})(w_i {\rm q_i} - \overline{w_i {\rm q_i}}) \right],
\end{equation}
\begin{equation}
C = \overline{{\rm q_i}}^2 \sum_{i=1}^N \left( w_i - \overline{w_i} \right)^2,
\end{equation}
{where \( \overline{w_i} \) is the mean weight and ${\rm q_i} = w_i \cos(2\theta_i)$.}

As a result of inhomogeneous observing strategies, and the presence of star clusters in some fields, some pixels have a high number of stars. 
%While most pixels have uncertainties of $0.5^\circ - 2^\circ$ in the mean polarization angle, some have uncertainties of $\sim 0.01^\circ$. To avoid assigning too high a weight to these pixels, 
{The weighted mean orientation in such pixels can have an uncertainty of less than $1^\circ$. However, the systematic uncertainty in the calibration of the polarization angle of stars is $\sim 1^\circ$ \citep{Clemens2012, Blinov2023}. To avoid underestimating the total uncertainty of the polarization angle, } we restrict the minimum uncertainty of the polarization angle in any pixel to  $\sigma_{\theta} = 1^\circ$. We have checked that changing this lower limit from $0.5^\circ - 2^\circ$ does not affect the results of the mean relative orientation of stellar polarization with respect to the RW.

\subsection{Statistics of angles}

In this work we wish to compare the polarization angle of stars to the projected shape of the RW on the plane of the sky. We quantify the significance of the alignment between two sets of angles with the Projected Rayleigh Statistic (PRS, \citealt{Jow2018}). The PRS is a measure of the narrowness of a distribution of angle differences. Values close to zero imply a random distribution. Values that are highly positive (negative) imply alignment (orthogonality). We compute the PRS, symbolized by $\rm V$, taking into account measurement uncertainties, as:
\begin{equation}
{\rm V} = \frac{1}{\sqrt{\sum_{i=1}^N w^2_i/2 }}\sum_{i=1}^N w_i \cos{2\Delta\psi_i} \; ,
\label{eqn:prs}
\end{equation}
where $\Delta\psi_i$ is the difference between two angles and $w_i$ is the weight as defined for Eq.~\ref{eqn:circmean}. 

As defined in Eq.~\ref{eqn:prs}, the value of the PRS depends on the number of measurements. To be able to reliably compare the PRS among datasets with different numbers of measurements, we normalize the PRS by its maximum possible value, $\rm V_{max}$, i.e. Eq.~\ref{eqn:prs} when all angles are zero:
\begin{equation}
{\rm \frac{V}{V_{max}}} = \frac{V}{\frac{1}{\sqrt{\sum_{i=1}^N w^2_i/2 }}\sum_{i=1}^N w_i}  \; .
\label{eqn:prs_normed}
\end{equation}

In Sect.~\ref{sec:results} we compare the normalized PRS, $\rm V/V_{max}$, for a set of angles of interest, to that of a uniform distribution of angle differences to quantify the significance of the alignment \citep{Jow2018}. 

%%%%%%%%%%%%%%%%%%%%%%%%%%%%%%%%%%%%%%%%%%%%%%%%%%

\begin{figure*}
\includegraphics[width=\textwidth]{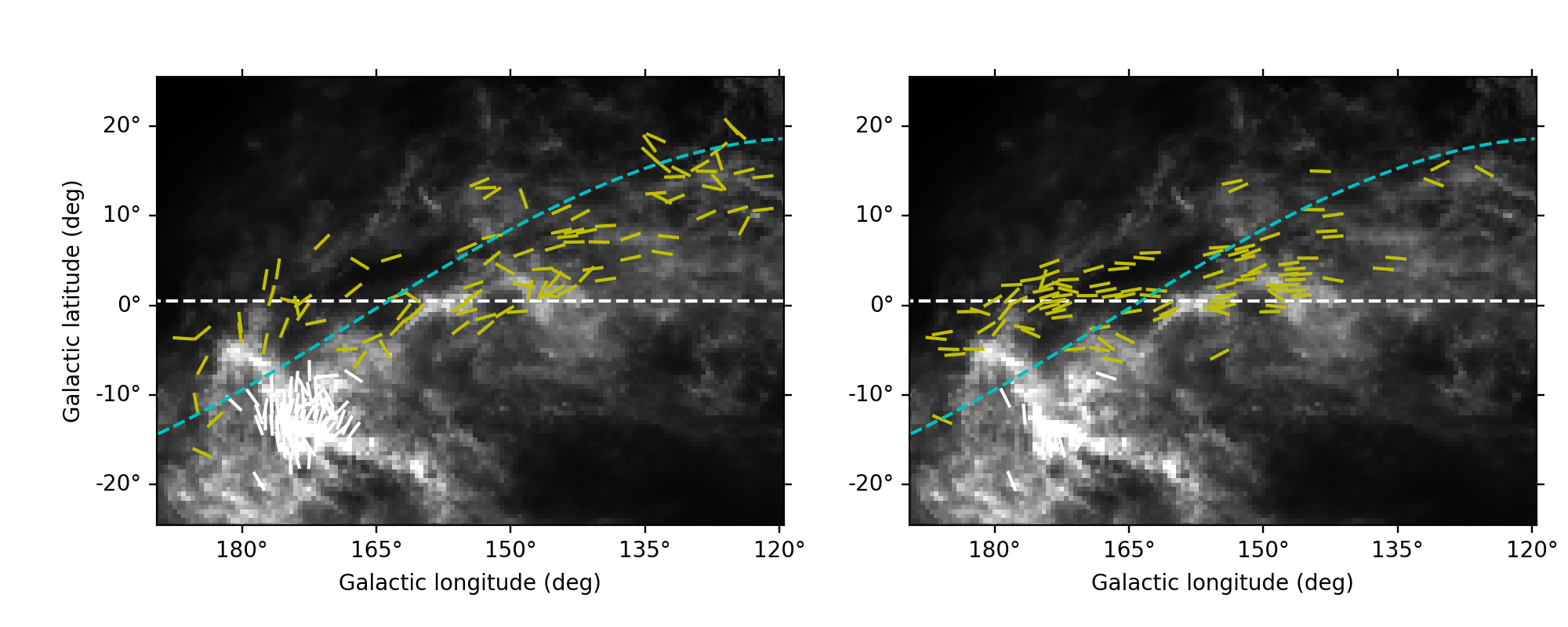}
\caption{Comparison between pixelized star polarization angle maps for stars within 400~pc (left) and stars beyond 1.2~kpc (right). Each yellow line segment traces the mean polarization angle of stars in a given pixel. The polarization angles of pixels towards the TMC (white segments) are excluded from the analysis. The Galactic plane is marked with a white horizontal dashed line. The background image shows the cumulative extinction out to 400~pc from \citet{Edenhofer} (as in Fig. \ref{fig:planck_stars}).}
\label{fig:segments_compare}
\end{figure*}

\section{Results}
\label{sec:results}

\begin{figure*}
\includegraphics[width=\textwidth]{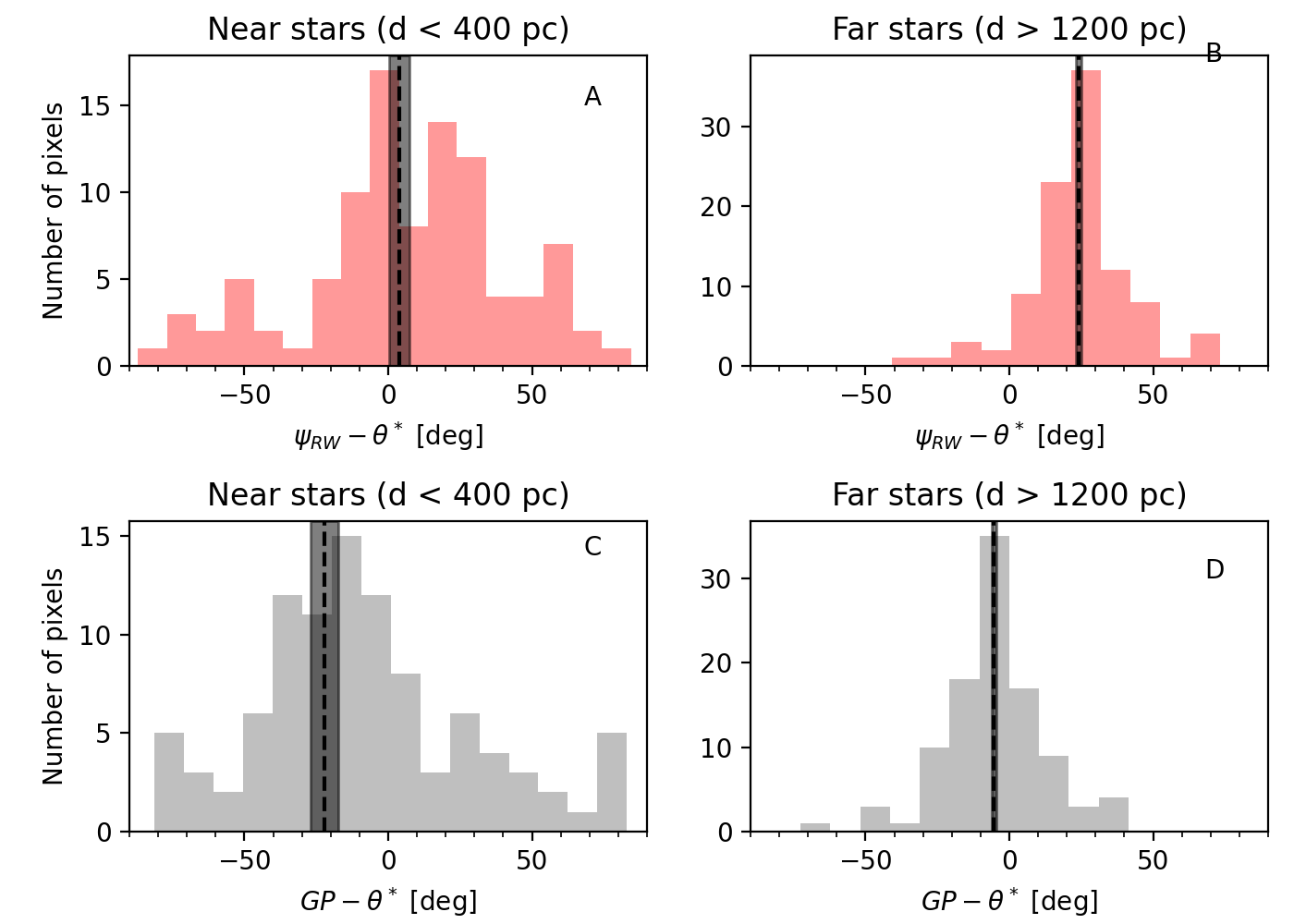}
\caption{Relative orientations of pixelized star polarization angles and the position angle of the RW (red, top row) or the Galactic midplane (gray, bottom row). Panels A and C show distributions for stars within 400~pc, while panels B and D those for stars beyond 1.2~kpc. The vertical gray band in each panel is centered on the circular mean of each distribution (dashed vertical line) and has a width equal to 2 times the uncertainty on that mean. }
\label{fig:galpa-hist}
\end{figure*}

\subsection{Magnetic field orientations towards the Radcliffe wave}
\label{sec:results1}

In Fig. \ref{fig:planck_stars} we compare the polarization angles of dust emission at 353~GHz from \textit{Planck} (after rotation by 90$^\circ$) with stars in the default near sample (stellar distances $d < 400 \; \rm pc$). We show the \textit{Planck} polarization data as segments (light brown lines), where the tilt of the segment corresponds to the angle $\theta_{dust}$ at that pixel, while the segment length is the same for all pixels. For ease of visualization, we show the \textit{Planck} data at $N_{\rm{side}} = 32$, while we use $N_{\rm{side}} = 64$ for the quantitative analysis. Each yellow line segment represents the mean polarization angle of stars in that pixel. The cyan line shows the projection of the RW model on the sky in this area, while the background image is the extinction out to 400~pc, encompassing the bulk of the extinction from the RW in the nearest-approach region. 

We observe a difference between these two tracers of the magnetic field. The \textit{Planck} data show a mean orientation of the magnetic field of -88$^\circ$; essentially parallel to the midplane of the Galaxy. The only significant local deviation is seen towards the TMC. The magnetic field there is known to be dominated by the TMC, and starlight polarization is well-aligned with the magnetic field traced by polarized dust emission \citep{Soler2016}. In contrast to the dust emission, the magnetic field as traced by stars within $d <$ 400 pc shows an offset with respect to the midplane, most notably around longitude $l = 165^\circ$.

We quantify this offset between the mean orientation traced by stars and by the dust emission as follows. We  construct pixelized maps of the stellar polarization angles at $N_{\rm{side}} = 64$ (resolution approximately $1^\circ$) including stars within 400 pc (see Sect.~\ref{sec:methods}). 
We select pixels where the polarization angle uncertainty is $< 12^\circ$ (corresponding to an S/N cut in polarization of 2.5), both in the \textit{Planck} map and in the stellar polarization map. This selection cut on the stellar polarization S/N only removes 2 of the 164 pixels, while 85\% of the pixels have uncertainties in the mean polarization angle of less than 5$^\circ$. The \textit{Planck} polarization data that remain are highly significant, with all pixels having S/N > 8. We exclude stars near the TMC. The distribution of angle differences between the \textit{Planck} $\theta_{dust}$ and the pixelized stellar polarization angles $\theta^*$ is shown in the inset of Fig.\ \ref{fig:planck_stars}. We observe that the distribution is offset from 0$^\circ$. The weighted circular mean of the distribution is -13$^\circ \pm$6$^\circ$. 
This offset reflects a shift in the mean orientation of the magnetic field as traced by the stars compared to the dust emission. The standard deviation of the distribution of angle differences is 38$^\circ$, much larger than the uncertainties of the mean polarization angle in individual pixels ($\sim 1^\circ$). These differences may arise from line-of-sight integration effects: dust unrelated to the RW exists at distances beyond 400 pc (Fig. \ref{fig:approach}), and likely contributes to the polarization of the dust emission traced by \textit{Planck}.

To investigate whether this offset occurs at a specific distance, we  compare the stellar data within 400~pc to the far sample (beyond 1.2~kpc). 
We construct maps of the weighted mean polarization angle within pixels of $N_{\rm{side}} = 64$, using the equations described in Sect.~\ref{sec:methods} and show them in Fig.~\ref{fig:segments_compare}. Data from stars in the near sample are shown in the left panel and data from stars in the far sample are shown in the right panel. At each pixel location, we show a yellow line segment representing the mean Galactic polarization angle of stars in that pixel. Sightlines towards the TMC are excluded from both samples - we show their corresponding Galactic polarization angles with white lines.  

In Fig.~\ref{fig:segments_compare}, the orientations of the  stellar polarization measurements of the near sample show significant offsets from those of the far stars. The polarization angles of the far sample star pixels are more aligned with the Galactic midplane, in agreement with the \textit{Planck} data. In contrast, the orientations of the magnetic field traced by nearby stars appear to be aligned with the projected position angle of the RW model (cyan line).

We quantify the relative orientations between the RW model and the stellar polarization data as follows. For each pixel in the binned star polarization map, we find the nearest (in projection) position of the RW model. We calculate the projected position angle of the model at that location ($\rm \psi_{RW}$). Then, we compute the angle difference between the mean star polarization angle in the pixel and  $\rm \psi_{RW}$. 

Figure~\ref{fig:galpa-hist} shows histograms of the relative orientations between the binned stellar polarization angles and the default RW model for the two aforementioned samples: stars in the near sample out to 400~pc (panel A) and stars in the far sample (panel B). The near sample distribution has a weighted circular mean of {4$^\circ\pm 4^\circ$}\footnote{{For all angle difference distributions we compute the error on the mean via the equations presented in Section \ref{sec:pixelization}}.}. The mean orientation of the polarization of these stars is consistent with alignment with the RW shape. 
In contrast, the far star sample distribution has a circular mean of {24$^\circ\pm 1^\circ$}, significantly offset from alignment with the RW. 

We also compare the polarization angles of star pixels with respect to the midplane of the Galaxy in the bottom panels of Fig.~\ref{fig:galpa-hist}. In this case, the far sample (panel D) shows a circular mean much closer to alignment: -{5}$^\circ\pm ${1}$^\circ$. The distribution of relative orientations for the near sample with respect to the midplane (panel C) has a circular mean of -20$^\circ \pm ${5}$^\circ$, inconsistent with midplane alignment.

We conclude that the near stars trace a plane-of-sky magnetic field that is aligned with the projected shape of the RW, while the far stars show polarization angles that are well aligned with the direction of the Galactic midplane.

The spread of the distribution of relative orientations also changes dramatically when considering the near vs. far samples. The standard deviation of the distribution of relative orientations for nearby stars with respect to the RW is {35}$^\circ$. The far sample has a standard deviation of 17$^\circ$. These values do not change appreciably between the midplane and RW comparisons. 

\begin{figure*}
\includegraphics[width=\textwidth]{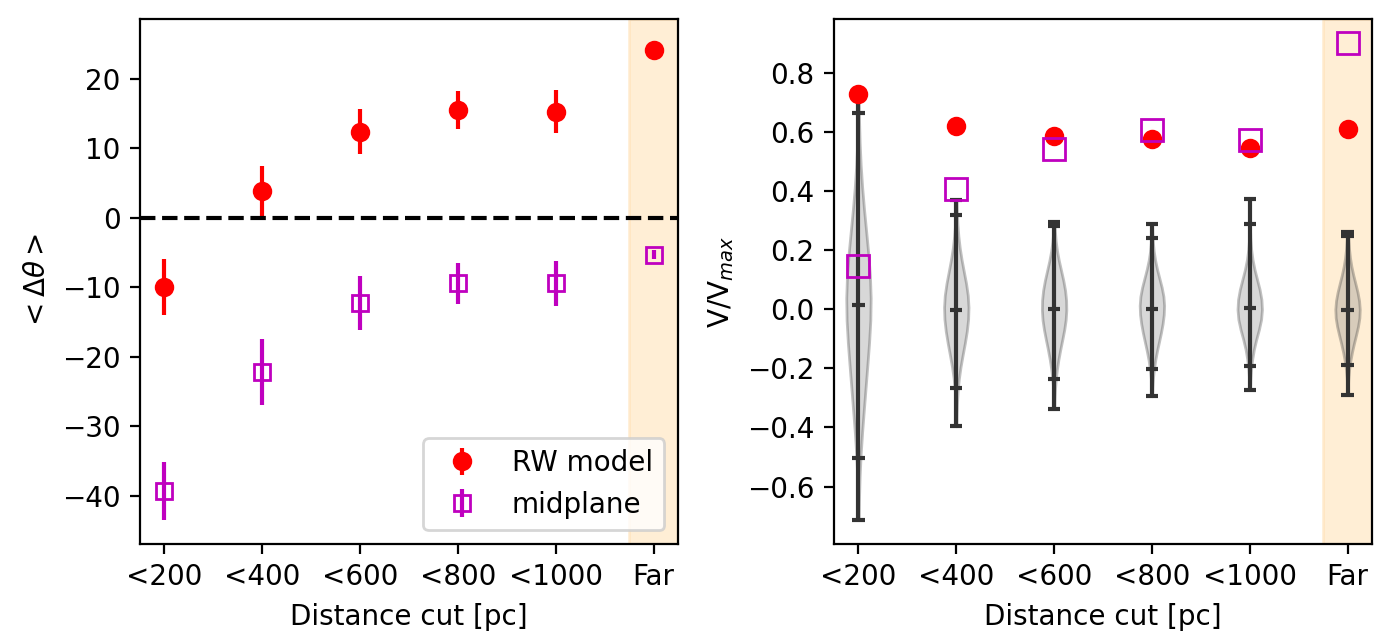}
\caption{Comparison of alignment and alignment significance as a function of the maximum distance of stars. Left: circular means ($\left<\Delta \theta\right>$). Right: normalized PRS ($V/V_{max}$). Each panel shows values of these quantities computed for (a) the relative orientation of star polarization angles compared to the RW (red circles), (b) the relative orientation with respect to the midplane (magenta squares). The violin shapes 
show the distribution of normalized PRS values obtained for a uniform distribution of angles for the same uncertainties as the corresponding stellar samples. Horizontal lines mark the median, 3-sigma lower and upper limits and the minimum and maximum value of each distribution.}% In all cases except the far distance sample, the alignment of stellar polarizations with the RW is more significant than with the midplane.}
\label{fig:violin}
\end{figure*}

One possible reason for the narrower spread in the distribution of angle differences for the far stars is that we are probing the magnetic field averaged over larger volumes. The linear size of the RW within the nearest approach region is 350~pc, while the minimum separation between pixels of 1$^\circ$ corresponds to 5~pc (taking the bulk of the dust to lie at a distance of 300~pc, Fig.~\ref{fig:approach}). The far stars trace the cumulative Stokes parameters out to large distances and therefore may exhibit less scatter due to %the 
averaging along the line of sight. In addition, as a result of our observing strategy, we have multiple far stars in each field of view. Therefore, star measurements are averaged in the plane of the sky during the pixelization process. For a distance of 3~kpc, stars averaged over $1^\circ$ pixels are tracing a magnetic field averaged over 50~pc projected linear size. In short, averaging both along the line of sight and across the plane of the sky for the far sample is likely the main reason for the much reduced scatter of the distributions of relative orientations for the far stars.

\subsection{Quantifying the relative orientations as a function of distance}
\label{sec:prs}

In the previous section, we have shown that the polarization angles of stars closer than 400~pc differ substantially from those of stars in the far sample. We have also shown that the distribution of relative orientations of starlight polarization  within 400~pc compared to the RW model peaks at $\approx$ 0$^\circ$, consistent with an alignment of the magnetic field traced by nearby stars with the shape of the RW as projected on the sky. In this section, we quantify the significance of this alignment between the magnetic field and the RW as a function of stellar distance. 

We construct samples of stars with different maximum distances, starting from 200~pc and incrementally increasing the maximum distance of the stars by 200~pc to a maximum of 1~kpc. The far sample remains as defined originally (minimum distance of 1.2~kpc).

To quantify the alignment between two sets of angles, we use two measures: i) the circular mean of the distribution of relative orientations and ii) the normalized PRS. The former quantifies the proximity of the mean relative orientation to zero (indicating alignment), while the latter quantifies the significance of that alignment compared to a uniform distribution (a measure of the spread of the distribution of relative orientations). We expect a significant alignment to manifest as both a near-zero mean relative orientation and a high normalized PRS (significant compared to a uniform distribution). 

Figure~\ref{fig:violin} (left) shows the circular mean of the distribution of relative orientations, $\left<\Delta\theta\right>$, of pixelated stellar polarization with respect to the RW model (red circles) and with respect to the Galactic plane (open square symbols), as a function of the maximum distance cut. The first sample extending out to 200~pc shows {an offset of $\sim 10^\circ$ from the RW orientation. This sample has the smallest number of pixels}. As the maximum distance is increased,  and the number of stars sampled increases, the error on the circular mean decreases. The samples out to 400 pc show a mean relative orientation with respect to the RW consistent with alignment within 1 $\sigma$. From the sample with < 600~pc onwards, we observe a non-negligible offset between the stellar polarization orientations and those of the RW (red circles). The largest offset with respect to the RW is seen for the far sample (right-most red circle). 

At the same time, the mean relative orientation of starlight polarizations compared to the Galactic plane is never consistent with 0$^\circ$. The far star sample has a mean orientation with the smallest offset compared to the Galactic plane of -7$^\circ$ (open square symbols). 
We conclude that for stars out to 400~pc, the polarizations are on average well-aligned with the RW, while this is not the case for alignment with the midplane direction. 

Next, we quantify the significance of the alignment discussed above. For each pixelized star sample, we compute the normalized PRS (Eq.~\ref{eqn:prs_normed}) and compare it to the normalized PRS of a uniform distribution. We randomize the relative orientation angles by drawing values from a random uniform distribution in the range $[-90^\circ, 90^\circ]$. For each measurement, we sample an `observation' by drawing from a Normal distribution centered on the random value obtained from the uniform distribution, with a standard deviation equal to the corresponding uncertainty in the measured polarization angle of the pixel. We repeat the generation of a set of random angles for each near and far sample 2000 times and compute the PRS each time.

Figure~\ref{fig:violin} (right) shows the normalized PRS for the  distributions of relative orientations at each distance limit: 
(a) the pixelized stellar polarization versus RW position angle ($\psi_{\rm{RW}}- \theta^*$, red circles), 
(b) the pixelized stellar polarization versus midplane direction ($GP - \theta^*$, purple squares),
and (c) the case of a randomized distribution of angle differences (violin shapes).

We find that the alignment with the RW seen for the near star sample $d<400\rm ~ pc$ is significant: the normalized PRS is greater than the values obtained from the randomized distributions. For samples with $d < 600 ~\rm pc$ or larger, though the PRS of the $\psi_{\rm{RW}} - \theta^*$ distributions are significant, the circular means are not consistent with alignment with the RW.

The offset between the RW shape and the stars beyond 600~pc may result from the presence of dust structures unassociated with the RW, some of which can be seen in Fig.~\ref{fig:approach} (right). 
The far sample shows a highly significant PRS of the %$90^\circ - \theta^*$, 
GP $- \theta^*$, and a small offset of the mean from 0$^\circ$, indicating that the far stars are more well-aligned with the midplane than with the RW, as seen initially in Fig.~\ref{fig:segments_compare}.

\begin{figure*}
\centering
\includegraphics[width=\textwidth]{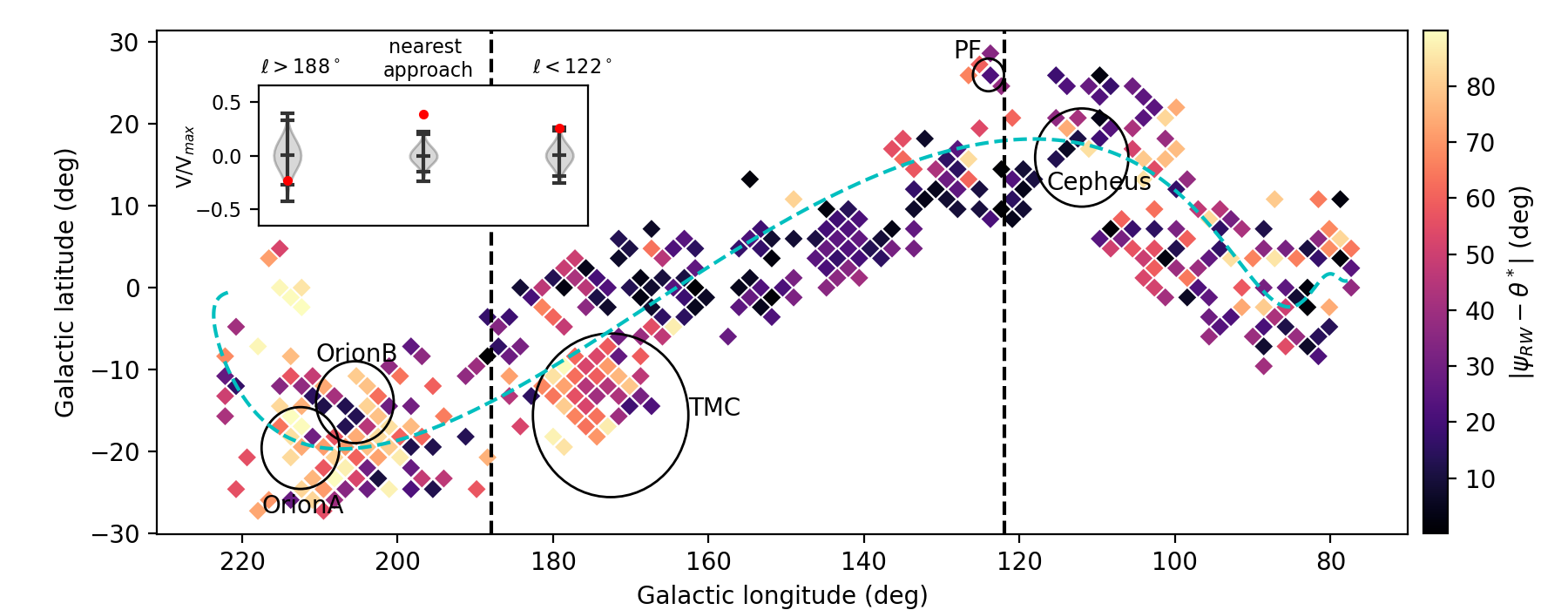}
\caption{Map of angle differences between the RW spine and pixelized stellar polarization angles. This covers the entire extent of the RW, for stars with distances $< 1.2$ kpc. The molecular clouds annotated are: TMC, Orion A, Orion B, Cepheus, and the Polaris Flare (PF).  Vertical dashed lines mark the nearest approach region within $l = (122^\circ, 188^\circ)$. The inset shows the normalized PRS for the three longitude ranges indicated in the main panel ($l > 188^\circ$, the nearest approach range, and $l < 122^\circ$). Violin plots and red markers in the inset are in accordance with Fig.\ \ref{fig:violin}.}
\label{fig:RW_diffs_all}
\end{figure*}

\subsection{Relative orientations spanning the entirety of the RW}
\label{sec:all_l}

In the previous sections, we focused on tracing the magnetic field of the RW at its nearest approach to the Sun. We showed that the RW magnetic field is not consistent with lying along the midplane of the Galaxy. Instead, the magnetic field appears to be aligned with the shape of the RW spine in projection. The next question pursued is whether this apparent alignment holds throughout the full extent of the RW.

In Fig.~\ref{fig:RW_diffs_all} we show the absolute relative orientations between the position angle of the RW and the pixelated stellar polarizations. The data used here include stars out to 1.2~kpc, the maximum distance of the RW. The stellar data have again been pixelized to $N_{\rm{side}} = 32$ for better visualization, but the results are consistent with those at $N_{\rm{side}} = 64$. We only show pixels with uncertainty in angle differences $< 12^\circ$. It appears that alignment over multiple adjacent pixels is only observed within the range of longitudes of the RW nearest approach ($l = 122^\circ$ - 188$^\circ$). %A smaller isolated area of alignment is seen towards $l = 110^\circ$. 

To quantify the degree of alignment of the RW magnetic field with the RW shape, we separate the data shown in Fig.\ \ref{fig:RW_diffs_all} into three longitude ranges: the range of the RW nearest approach, the range of RW longitudes $l > 188^\circ$, and that with $l < 122^\circ$. In the inset of Fig.\ \ref{fig:RW_diffs_all}, we show the normalized PRS for the three longitude ranges, with symbols as in Fig.\ \ref{fig:violin}. As can be seen from comparing with Fig.\ \ref{fig:approach}, the RW is oriented mostly along the line-of-sight in the two extreme longitude ranges, and mostly along the plane-of-sky in the middle one. We see that only the longitude range of nearest approach has a significant normalized PRS, while outside this range the PRS values are consistent with arising from random distributions. 
We speculate three possible reasons for this.

The first possibility is that the large-scale Galactic magnetic field is not aligned with the spine of the RW, with the exception of the nearest approach region. 
The second possibility, is that local small-scale distortions of the magnetic field are dominating the observed polarization angles of the stars. 
We note that Fig.~\ref{fig:RW_diffs_all} and the corresponding PRS analysis includes stars tracing various molecular clouds that appear along the RW (notably, the TMC, Orion~A and~B as well as the Polaris Flare). We have annotated their locations and approximate sizes as circles in the figure. The central positions are taken from \citet{Zucker2021_clouds}. For the Polaris Flare we used the center of the \textit{Herschel} map \citep[e.g.][]{Panopoulou2016}. The diameter of each circle corresponds to the projection of the maximum extent of the cloud's skeleton defined in \citep{Zucker2021_clouds}. Feedback events, gravitational collapse and turbulence may have distorted the magnetic field from its initial configuration in such dense molecular clouds. The magnetic field around the Orion clouds is known to have been affected by the Orion-Eridanus superbubble \citep{Soler2018}.
The third possibility is that the large-scale magnetic field is aligned with the RW in 3D, but the alignment is lost in projection as the structure moves away from the plane of sky and becomes increasingly parallel to the line-of-sight. In this case we predict greater scatter in the polarization angles of the stars in the regions where the RW is pointing mostly along the LOS.
Given the evidence in the literature that the magnetic field is parallel to the Local Arm in 3D (see Sect. \ref{sec:discussion}), we favor the latter possibility. More detailed modeling of the RW magnetic field geometry is needed to distinguish among the above scenarios.

\subsection{Dependence on the choice of RW model}

We repeat the analysis presented in Sect. \ref{sec:results1}, \ref{sec:prs}, and \ref{sec:all_l} for the 20 alternative models of the RW spine (Sect. \ref{sec:data}) to investigate any dependence of the results on the choice of model. We find that the star sample within 400 pc is aligned with the RW for 13 of the 20 models. In three out of the 20 models the sample within 600 pc also shows alignment with the RW. Models that cross the midplane at longitude less than 159$^\circ$ have $\left<\Delta \theta\right>$ that are inconsistent with alignment. This indicates that the latter models do not provide a good description for the mean magnetic field in the RW.

The results of Figure \ref{fig:RW_diffs_all} are robust to the choice of model. For all models the PRS in the nearest approach longitude range is significant, while that in the other two longitude ranges remains consistent with that from a random distribution of angles.

\section{Discussion}
\label{sec:discussion}

We have performed a study of the polarization angles of stars towards the region of the sky where the RW is nearest to the Sun. We have determined that the polarization angles vary with distance in a systematic fashion. Stars within 400~pc of the RW trace a magnetic field that is aligned with the projected shape of the RW. In contrast, stars farther than 1.2~kpc have polarization angles that are preferentially aligned with the Galactic plane (Fig.\ \ref{fig:violin}). 

We note that the observed polarization angles of stars correspond to the cumulative polarization tracing dusty structures out to the distance of each star. At the distance of nearest approach of the RW, the stellar polarization is dominated by dust in the RW itself, thus tracing the local-to-the-RW magnetic field orientation (Fig.\ \ref{fig:approach}). For distances beyond the RW, stellar polarization may arise from multiple components along the line of sight, similarly to what is found for polarized dust emission throughout the sky \citep[e.g][]{Halal2024}. A tomographic decomposition of the Stokes parameters with distance would be necessary to determine whether the magnetic field is aligned locally with the midplane at  distances beyond the RW (\citealt{Pavel2014}; \citealt{Panopoulou2019}; \citealt{Pelgrims2023}, \citeyear{Pelgrims2024}, \citealt{Doi2024}).

Previous analyses of stellar polarization towards the Local Arm found a mean polarization orientation parallel to the Galactic midplane \citep{Fosalba2002}, especially for stars farther than 1~kpc \citep{Heiles1996}. In our study we find that the magnetic field shows a mean offset from the midplane of $18^\circ$ over the nearest approach longitude range. At the longitude where the RW crosses the midplane, both the RW and the magnetic field form an angle of $\sim 30^\circ$ with the plane. The linear size of the region within which the magnetic field departs from plane-parallel geometry is 350 pc. 

Localized, smaller angular-scale deviations from a parallel to the midplane geometry have been noted towards other regions of the Galaxy (\citealt{Pavel2014}; \citealt{Versteeg2023}; \citealt{Doi2024}). \citet{Clemens2020} pointed out that deviations from the Galactic midplane direction are relatively common when looking at the observed polarization angles of stars in their extensive H-band polarization survey towards the inner Galaxy. 
Determining whether the large-scale deviation found in the RW is an exception, or whether such deviations are more prevalent throughout the disk, would be essential for an accurate description of the large-scale Galactic Magnetic Field (GMF). 

\subsection{Implications for large-scale GMF modeling.}
\label{sec:GMF}

Determining the 3D geometry of the coherent component of the magnetic field in the Local Arm is necessary for constructing accurate models of the GMF. The RW traces only a 3-kpc-long section of the Local Arm, while the entire arm extends over 8 kpc as determined by maser observations \citep{Reid_2019}.

Previous studies using stellar polarization or rotation measures of pulsars have shown that the coherent component of the magnetic field in the Solar vicinity points towards longitude $l = 70 - 95^\circ$ (\citealt{Manchester1974}; \citealt{Rand1989}; \citealt{Heiles1996}).
Significant discrepancies were initially found between stellar polarization and rotation measures (\citealt{Ellis1978}; \citealt{Inoue1981}). However, later estimates of the local direction of the magnetic field in the dusty ISM confirm a longitude range of $l = 72 - 85^\circ$ from stellar polarization, (\citealt{Heiles1996}), and $l = 70^\circ - 77^\circ$ at high latitudes from polarized dust emission (\citealt{Planck_Galactic_cap_2016}; \citealt{Pelgrims2020}). This longitude range was known to correspond to the direction in which we observe the Local Arm end-on (e.g., \citealt{Heiles1996}), and also corresponds to the end-point of the RW (Fig.~\ref{fig:RW_diffs_all}). 

These previous studies inferred that the magnetic field is aligned with the Local Arm - a conclusion which is also confirmed by Faraday rotation towards extragalactic sources \citep{Hutschenreuter2020}. On the basis of stellar polarization data, the 3D orientation of the GMF in the Local Arm was determined by modeling the stellar polarization fractions as a function of longitude (\citealt{Inoue1981}; \citealt{Heiles1996}). In Sect.~\ref{sec:results}, we investigated a complementary tracer of the 3D direction of the magnetic field: the relative orientation of polarization angles with respect to the RW as a function of longitude. If the magnetic field was aligned with the RW throughout its extent, we would qualitatively expect near-perfect alignment in projection in the region where the RW is observed entirely in the plane of the sky. Conversely, due to distortions of the magnetic field, we would expect a loss of alignment in the projected relative orientations for the directions in which the RW is viewed end-on. These expectations qualitatively match the observed relative orientations in Fig.~\ref{fig:RW_diffs_all}. A more complete stellar sample and detailed modeling are needed to infer whether 3D alignment with the RW is the best-fit geometry of the large-scale magnetic field of the structure.

Current GMF models constrain the magnetic field geometry to lie along the spiral arms as determined by a model for the thermal electrons (\citealt{JanssonFarrar2012,Jaffe2019}).
The RW is the gas reservoir of the Local Arm in the Solar vicinity, and its shape is found to deviate from traditional models of spiral arms \citep{Zucker2023}. Our observations provide insights into the magnetic field in the dusty, neutral phase of the ISM. If the magnetic field is indeed found to follow the RW perturbation in 3D, then GMF models must be updated to include this sinusoidal perturbation in the large-scale magnetic field. It would be interesting to determine whether other observed corrugations in the gaseous disk (e.g., \citealt{Veena2021}) also have a counterpart in the GMF geometry.

Finally, our constraints on the GMF geometry towards the RW have implications for the distance determination of a prominent feature in the radio sky known as the Fan region \citep{Brouw1976}. 
The distance to the Fan region remains unclear. Depolarization by distant ionized sources implies that $~30-40\%$ of the emission at 1.5 GHz arises from a distance larger than $ 2~\rm kpc$ \citep{Hill2017}. However, recent modeling of the polarized synchrotron emission suggested a local origin, associated with the RW \citep{West2021}. The polarization angles of the synchrotron emission in the Fan region trace a magnetic field that is parallel to the midplane. Given our findings that the magnetic field is not parallel to the midplane at the distance of the RW, it is unlikely that the Fan region coincides with the RW as suggested by \citealt{West2021}. Our results, however, do not rule out the existence of a component of the cold, dusty ISM that coincides with the synchrotron emitting volume giving rise to the Fan region beyond the distance of the RW.

\subsection{Implications for the formation mechanism of the RW}

Three classes of models have been proposed to explain the formation of the RW. The first model proposes that the RW arose from a perturbation of the Galactic disk by the passage of a dwarf Galaxy (\citealt{Alves2020}).
This scenario addresses the fact that perturbations are observed in the kinematics of stars (\citealt{Thula_2022}). A second scenario posits the observed undulation of the RW is the result of feedback events (e.g. multiple clustered supernovae at different locations along the RW, \citealt{Alves2020}). 
A third scenario is that the RW is the result of an instability inherent in the disk, such as a Kelvin-Helmholtz instability (\citealt{Fleck2020}). %The feedback scenario is disfavored based on fine-tuning arguments (\citealt{Konietzka2024}), while the Kelvin-Helmholtz instability would not explain the disturbance of the stellar disk.

Our magnetic field observations provide additional constraints that any viable mechanism should satisfy. The magnetic field is ordered over lengthscales of 300-400 pc and exhibits an inclined crossing with respect to the midplane, at the location where the RW crosses the midplane. In projection, the magnetic field appears aligned with the RW spine. We hypothesize that the magnetic field is aligned in 3D with the RW, as suggested by our results in Fig. \ref{fig:RW_diffs_all} and the previous discussion. It is possible that the aforementioned scenarios would predict different magnetic field geometries, e.g. depending on the level of turbulence they induce in the gas as a function of scale. Explicit predictions for the magnetic field from these types of formation mechanisms would require magneto-hydrodynamical simulations. 

A potentially interesting consequence of the presence of a large-scale perturbation in the magnetic field is whether the Parker instability would be triggered (\citealt{Parker1966}). 
If we hypothesize that the magnetic field lies parallel to the RW throughout its length, then the magnetic field would exhibit an oscillatory pattern, reminiscent of this instability. 
The spacing between peaks in the damped sinusoid model describing the RW spine is $\sim$~2~kpc, comparable to the wavelength of the Parker instability parallel to the magnetic field (1 - 2~kpc, \citealt{Heintz2020}; \citealt{Tharakkal2023}). 
The structure of the magnetic field, which exhibits a coherent component over 350~pc (at least) and crosses the midplane is reminiscent of the antisymmetric mode observed in simulations of the Parker instability (\citealt{Mouschovias2009}).
The timescale for the Parker instability to grow from a small perturbation is $\sim$ 100~Myr in Solar neighborhood conditions (\citealt{Heintz2020}). However, the growth of the instability could be much faster for large perturbations (e.g. due to the passage of a spiral shock wave, \citealt{Mouschovias1974}). For example, \citet{Habegger2023} found that a point-like injection of a significant energy from cosmic-rays (from a cluster of supernovae explosions) can reshape the magnetic field and the ISM within a timescale as short as 20~Myr. If the passage of a dwarf galaxy has indeed caused an initial perturbation, it remains to be shown whether the instability would be excited in the disk.

The Parker instability has been difficult to robustly observe and it remains unclear whether it is suppressed in galaxies like the Milky Way \citep{Kim2001,Kim2002,Tharakkal2023,Hopkins2024}. While its signatures are suggested in the Faraday rotation patterns of nearby galaxies (\citealt{Beck2015}), pinpointing its presence in the Galaxy has proven elusive. Observations of the instability in the Milky Way have been claimed in various works (\citealt{Vrba1977}; \citealt{Fukui2006}; \citealt{Sofue2017}), but are hampered by confusion effects due to the unknown 3D geometry of the magnetic field. It has been suggested that the instability in conjunction with supernova feedback is responsible for the predominance of vertical \HI~ filaments towards the inner Galaxy (\citealt{Soler2022}). If triggered by some initial perturbation related to the formation of the RW, the instability could grow to further affect the distribution of gas and magnetic fields in the structure over time. It would be interesting to investigate this possibility with magneto-hydro-dynamical simulations.

\section{Summary}
\label{sec:conclusions}

We have carried out an investigation of the magnetic field geometry of the Radcliffe Wave. We have combined archival stellar polarimetry with new NIR polarization measurements towards the nearest portion of the RW to trace the plane-of-sky component of the magnetic field. 

We have shown that the RW is the main dust structure along the line-of-sight for most sightlines within 1.2~kpc of the Sun (for sightlines within $10^\circ$ of the RW spine). As a result, the observed polarization angles of stars immediately background to the RW appear to trace the magnetic field of this structure. By isolating stars within 400~pc of the Sun, we find a significant departure (20$^\circ$) of the magnetic field from the midplane of the Galaxy. The plane-of-sky magnetic field appears aligned with the shape of the RW spine orientation within the longitude range $l \in [122^\circ,\, 188^\circ]$. In contrast, stars beyond 1.2~kpc have polarization angles preferentially aligned with the Galactic midplane, consistent with measurements of polarized dust emission by \textit{Planck}.

We have investigated the significance of the observed alignment of stellar polarizations with the RW as a function of distance. We have shown that the alignment of the magnetic field with the RW is most significant for stars within 400~pc of the Sun.

We have compared the relative orientation of stellar polarization and the projected geometry of the RW over its entire extent. Significant alignment between the two geometries (in projection) is best found in the sky region where the RW is at nearest approach to the Sun. This observation is consistent with the magnetic field geometry being aligned with the RW in 3D. We have discussed the implications of our findings for Galactic magnetic field models as well as possible formation scenarios for the RW itself.

%%%%%%%%%%%%%%%%%%%%%%%%%%%%%%%%%%%%%%%%%%%%%%%%%%

\section*{Acknowledgements}

We thank Ralf Konietzka for helpful discussions. The authors acknowledge Interstellar Institute's program "II6" and the Paris-Saclay University's Institut Pascal for hosting discussions that nourished the development of the ideas behind this work.
This study was based on observations using the 1.8 m Perkins Telescope Observatory (PTO) in Arizona, owned and operated by Boston University. Data were obtained using the Mimir instrument, jointly developed at Boston University and Lowell Observatory and supported by NASA, NSF, and the W.M. Keck Foundation. This study was partially supported by grant AST 18-14531 from NSF/MPS to Boston University.
VP acknowledges funding from a Marie Curie Action of the European Union (grant agreement No. 101107047). S.E.C. acknowledges support from the National Science Foundation under grant No. AST-2106607 and an Alfred P. Sloan Research Fellowship. 
JDS acknowledges funding from the European Research Council (ERC) via the Synergy Grant ``ECOGAL -- Understanding our Galactic ecosystem: From the disk of the Milky Way to the formation sites of stars and planets'' (project ID 855130). JA acknowledges funding from the European Research Council (ERC) via the Advanced Grant "ISM-FLOW" (101055318). JBT acknowledges support from the DFG via SFB1491 "Cosmic Interacting Matters" (project no.\ 445052434).
This work uses results from the European Space Agency (ESA) space mission \textit{Gaia}. \textit{Gaia} data are being processed by the \textit{Gaia} Data Processing and Analysis Consortium (DPAC). Funding for the DPAC is provided by national institutions, in particular the institutions participating in the \textit{Gaia} MultiLateral Agreement (MLA). The Gaia mission website is \url{https://www.cosmos.esa.int/gaia}. 
The work is based on observations obtained with \textit{Planck} (\url{http://www.esa.int/Planck}), an ESA science mission with instruments and contributions directly funded by ESA Member States, NASA, and Canada.
We make use of the HEALPix \citep{Gorski2005}, astropy \citep{astropy2022}, matplotlib \citep{Hunter:2007}, healpy \citep{Zonca2019}, and scipy \citep{2020SciPy-NMeth} packages. 

%%%%%%%%%%%%%%%%%%%%%%%%%%%%%%%%%%%%%%%%%%%%%%%%%%



\bibliographystyle{aa}
\bibliography{main} % if your bibtex file is called main.bib

%%%%%%%%%%%%%%%%%%%%%%%%%%%%%%%%%%%%%%%%%%%%%%%%%%

%%%%%%%%%%%%%%%%% APPENDICES %%%%%%%%%%%%%%%%%%%%%

%%%%%%%%%%%%%%%%%%%%%%%%%%%%%%%%%%%%%%%%%%%%%%%%%%

\label{lastpage}
\end{document}